\documentclass[aps,prb,preprint,amsmath,amssymb]{revtex4}
\usepackage{graphicx,subfigure, times}
\begin{document}

\title{Thermodynamic stability of oxygen point defects in cubic Zirconia}

\author{Amit Samanta}
\email{asamanta@princeton.edu}
\affiliation{$^1$Program in Applied and Computational Mathematics,
 Princeton University, Princeton, New Jersey 08544, USA}

\author{Thomas Lenosky and Ju Li}
\affiliation{$^2$Department of Materials Science and Engineering, 
 University of Pennsylvania, Philadelphia, Pennsylvania 19103, USA}

\date{\today}

\begin{abstract}

 Zirconia (ZrO$_{2}$) is an important material with technological applications
 which are affected by point defect physics. Ab-initio calculations are performed
 to understand the structural and electronic properties of oxygen vacancies and
 interstitials in different charge states in cubic zirconia. We find oxygen
 interstitials in cubic ZrO$_{2}$ can have five different configurations -
 $\langle 110\rangle$ dumbbell, $\langle 100\rangle$ dumbbell, $\langle 100\rangle$
 crowd-ion, octahedral, and $\langle 111\rangle$ distorted dumbbell. For a neutral
 and singly charged oxygen interstitial, the lowest energy configuration is the
 $\langle 110\rangle$ dumbbell, while for a doubly charged oxygen interstitial
 the octahedral site is energetically the most favorable. Both the oxygen interstitial
 and the oxygen vacancy are negative-$U$, so that the singly charged defects
 are unstable at any Fermi level. The thermodynamic stability of these defects
 are studied in terms of Fermi level, oxygen partial pressure and temperature. A
 method to determine the chemical potential of the system as a function of
 temperature and pressure is proposed. 
\end{abstract}


\maketitle
\section*{1. Introduction}
 
 Zirconia (ZrO$_{2}$) exists in three closely related phases with monoclinic,
 tetragonal and cubic symmetry. The monoclinic phase is thermodynamically stable
 near room temperature and up to 1400 K, while at higher temperatures tetragonal
 and then cubic phases become stable.$\cite{ZhaoV02}$ Cubic zirconia is stable
 above 2570 K.$\cite{ZhaoV02}$ However, the cubic phase can be stabilized at much
 lower temperatures by doping ZrO$_{2}$ with divalent or trivalent cations like
 Y$^{3+}$ or Ca$^{2+}$.$\cite{StapperBNP99}$ This cubic stabilized zirconia (CBZ) has high thermal
 shock resistance,
 high strength and toughness and has been under extensive investigation in recent
 years.$\cite{BogicevicW03,StapperBNP99}$ CBZ is also important for oxygen sensors
 and fuel cells.$\cite{Fleming77,DocquierC02,Eichler01}$ To satisfy charge neutrality
 condition after cation doping, these materials contain oxygen vacancies and interstitials
 whose concentrations are related to the dopant concentration.
 
 Unstabilized cubic zirconia is also an interesting material for high temperature
 applications. A proposed methodology for exterior coatings in high speed aircraft
 involves coating with ZrB$_{2}$/SiC ceramic.$\cite{OpekaTZ04, MonteverdeB03}$ Upon
 exposure to high temperatures in an oxidizing atmosphere, ZrB$_{2}$/SiC oxidizes to
 ZrO$_{2}$ and SiO$_{2}$+B$_{2}$O$_{3}$, a vitreous borosilicate glass which forms
 a sticky layer on the top.$\cite{BongiornoFKLMOTVY06}$ Little is known about point
 defects and their stability for the high temperature phase of ZrO$_{2}$, yet defect
 diffusion within the oxide scale is potentially a key issue for optimization of this
 material system, and experimental studies are difficult because of the extreme
 operating conditions.

 Moreover, the semiconductor industry has recently begun using hafnia and zirconia as a substitute
 for silicon dioxide in MOSFET gate dielectrics, an application for which they have merit because
 of their high dielectric constant.$\cite{FosterSGSN01,FiorentiniG02,RobertsonXF05}$ For this
 application the crystalline structure is generally the low-temperature monoclinic phase. Also CBZ
 has good resistance to radiation induced damage, leading to a number of proposed and actual nuclear
 applications.$\cite{GrynszpanBAMSVB07,ThomeGJGT07,ThomeGJGT06,DegueldreH03}$ For both of these
 applications point defect physics are crucial. In the gate oxide application point defects can
 trap and scatter charge carriers and create other unfavorable effects. In the nuclear applications
 the rapid recombination of radiation-induced defects gives zirconia its radiation resistance.

 In this paper we study point defect formation in cubic ZrO$_{2}$
 by using first principles density functional theory calculations.
 First we obtain the atomic structure and formation energy of the
 different intrinsic defects. Then we present results in which we
 calculate defect concentrations as a function of Fermi level,
 oxygen partial pressure and temperature, by studying the
 thermodynamic equilibria for the various defects with each
 other, with ambient oxygen, and with charge carriers. 

\section*{2. Method of calculation}

 We perform calculations using the Vienna Ab-Initio Simulation
 Package (VASP),$\cite {KresseF96, KresseFJ96, KresseJ99}$ a plane
 wave basis density functional theory (DFT) code, with generalized
 gradient approximation (GGA) and the exchange correlation potential
 of Perdew, Burke and Wang (PBE).$\cite{PerdewBE96}$ Spin polarization
 is taken into account in these calculations. The wave functions
 are expanded to a plane wave cut-off of 520 eV. Static relaxations
 are performed till the forces on individual atoms are smaller than
 10$^{-4}$ eV/$\rm{\AA}$ and total energy convergence within 10$^{-5}$
 eV is achieved. Initial ${\bf k}$-point sampling is performed with 1 to 20
 irreducible kpoints in the first Brillouin zone. The equilibrium
 lattice parameter of cubic ZrO$_{2}$ is 5.1276
 $\rm{\AA}$ (with 20 irreducible kpoints) which is close to
 previous density functional calculations and experimental
 values.$\cite{JomardPPMKH99, FosterSGSN01}$ All subsequent defect calculations are
 performed at this lattice constant, with no relaxation of the super
 cell unless otherwise mentioned. Total energy of a cubic ZrO$_{2}$
 primitive cell converged to within 6 meV with 4 irreducible
 kpoints. Hence, we use 4 irreducible points for all subsequent
 calculations.
 
 The band gap is given by the energy difference between the highest occupied electronic
 energy level and the lowest unoccupied energy level. We obtain a band gap of 3.26 eV
 which is smaller than 5.40 eV obtained using GW
 calculations.$\cite{FosterSGSN01, RobertsonXF05, JiangGRS10}$ DFT calculations are
 known to underestimate the band gap.$\cite{RobertsonXF05}$ To describe the DOS of the
 defect-free cell obtained from VASP calculations, a rigid shift of the conduction band
 edge is performed to obtain the experimentally reported band gap. The defect levels
 are not scaled while shifting the conduction band edge.$\cite{FosterSGSN01,RobertsonXF05}$ 

 Understanding the thermodynamic stability of charged defects involves incorporation or
 removal of electrons from the defect-free materials surrounding the defect. In the DFT
 calculations, the electronic charge neutrality of the supercell is achieved by a uniform
 jellium of equal and opposite charge. This results in long range interaction between the
 defect and its periodic images in the calculation which must be subtracted off from the
 total energy of the supercell $E_{\rm vasp}$. Thus $E$ = $E_{\rm vasp}$-$E_{\rm cor}$ is
 the corrected energy of the system. $E_{\rm vasp}$ is the energy obtained from VASP
 calculations for a sample containing defects. For a single defect the correction term
 is$\cite{MakovP95}$
\begin{equation}
E_{\rm cor}=-\frac{q^{2}\alpha}{4\pi L\kappa\epsilon_{\rm o}}
\end{equation}
where $q$ is the charge on the defect, $\alpha=5.04$ is the Madelung constant, $L$ is the
 length of the cubic supercell used for calculations, $\kappa$ is the
 dielectric constant of cubic ZrO$_{2}$ and $\epsilon_{\rm o}$ is the permittivity of free
 space. For cubic zirconia, we take $\kappa=36.8$.$\cite{ZhaoV02}$ Typical Madelung
 corrections in our supercells are of the order of 0.05$q^2$ eV, where $q$ is the charge of
 the defect (...,-2,-1,0,1,2,...).

 Given our primary interest in the high temperature cubic phase we calculate formation free
 energies for the lattice defects in zirconia as a function of temperature and oxygen partial
 pressure.

\section*{3. Defect geometry, electronic charge state and stability}
 Stability of point defects have to be analyzed with respect to some reference state. Since
 cubic zirconia is a high temperature phase and frequently used in contact with atmosphere
 we use oxygen gas molecule as the reference state for calculations of formation energies
 of oxygen vacancies and oxygen interstitials. 

\subsection*{3A. Reference state calculations}
 First-principles calculations of an oxygen molecule are performed at
 $T$ = 0 K. The triplet state is found to be the ground state of O$_{2}$. We find
 $E^{\rm vasp}_{\rm O_{2}}$=-9.86 eV and a bond length of 1.23 $\rm{\AA}$. The singlet
 state is 1.06 eV higher than the triplet state.  

\begin{figure}[thbp]
   \centering
   \includegraphics[width=0.5\textheight]{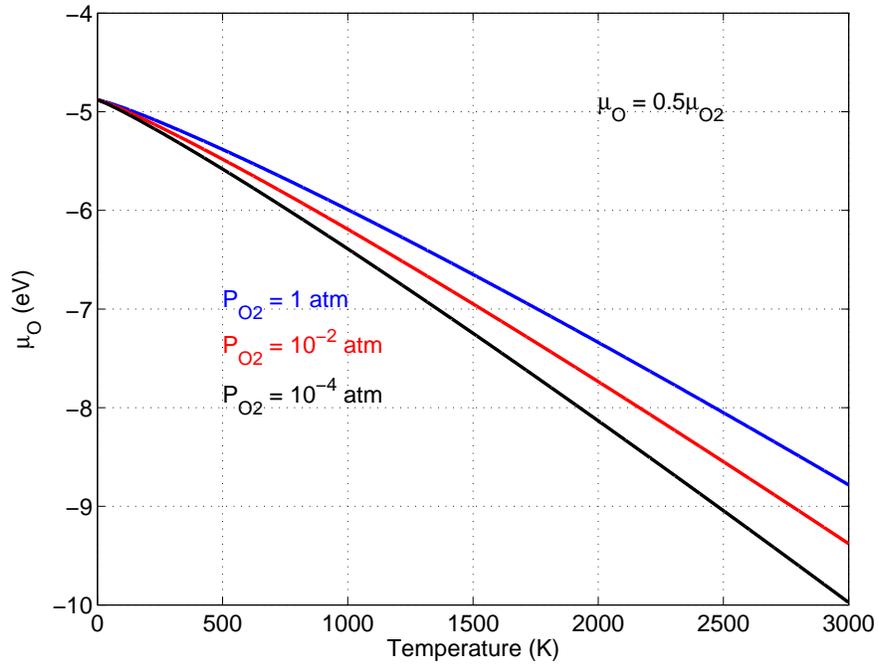}            
   \caption{ The chemical potential of an oxygen atom as a function of temperature at
     different oxygen partial pressures calculated from the quantum formulation of an ideal
     gas. $\left(\mu_{\rm O} = 0.5\mu_{\rm O_{2}}\right)$}
   \label{ChemOxy}
 \end{figure}

 The chemical potential ($\mu_{O_{2}}$) of an oxygen gas molecule changes with
 temperature as well as the oxygen partial pressure. An estimate of the free energy of
 an oxygen molecule can be made based upon the quantum formulation of an ideal gas. The
 free energy of an ideal diatomic gas has contributions from the translational free
 energy, vibrational free energy and rotational free energy.$\cite{McQuarrieDA}$ So the
 chemical potential of an oxygen molecule as a function of temperature $\left(T\right)$
 and pressure $\left(P\right)$ is given by :$\cite{McQuarrieDA}$ 
  \begin{equation}
   \mu = k_{\rm B}T\log\left(\frac{P\lambda^{3}}{k_{\rm B}T}\right) -
   k_{\rm B}T\log\left(\frac{4\pi^{2}Ik_{\rm B}T}{h^{2}}\right) +
   \frac{1}{2}\hbar\omega_{\rm o} +
   k_{\rm B}T\log\left[1-\exp\left(-\frac{\hbar\omega_{\rm o}}{k_{\rm B}T}\right)\right]
 \end{equation}
  where, $\lambda$ is the thermal de-Broglie wavelength, $k_{\rm B}$ is the Boltzmann constant,
 $h$ is Planck's constant and $I$ is the moment of inertia of the oxygen molecule.
 The chemical potential can also be expressed in terms of a reference pressure $P_{\rm r}$:
 \begin{equation}
   \mu = \mu_{\rm r} + k_{\rm B}T\log\frac{P}{P_{\rm r}}
 \end{equation}
 where, $\mu_{\rm r}$ is the chemical potential at reference pressure. The reference
 pressure is generally taken as 1 atm.
 From ab-initio calculations, we find the vibrational frequency of O$_{2}$
 is $\omega_{\rm o}$ = 1568.06 cm$^{-1}$, which is close to experimental result of
 $\omega_{\rm o}$ =1565.4 cm$^{-1}$.$\cite{Villars29}$ For our subsequent analysis,
 the chemical potential of an oxygen atom is taken as half of $\mu_{\rm O_{2}}$ :
 \begin{equation}
   \mu_{\rm O}\left(T,P\right) = \frac{1}{2}\mu_{\rm O_{2}}\left(T,P\right)
 \end{equation}
 Fig. $\ref{ChemOxy}$ shows the variation in the oxygen chemical potential with
 temperature at different oxygen partial pressures. From this analysis we obtain
 $\mu_{\rm O}$($T$ = 300 K, $P$ = 1atm) = -5.16 eV and $\mu_{\rm O}$($T$ = 2600 K,
 $P$ = 1atm) = -8.20 eV. Since, the cubic phase in undoped ZrO$_{2}$ is stable above
 2570 K, we have used $T$ = 2600 K, for subsequent analysis.

\subsection*{3B. Oxygen Vacancy}
 Oxygen vacancies in cubic ZrO$_{2}$ have been discussed previously because their
 motion is related to the ionic conductivity of zirconia. They have also been studied to
 understand the defect levels, which are important for applications in the semiconductor
 industry.$\cite{FosterSGSN01,RobertsonXF05}$
 In the low temperature monoclinic phase there are two
 non-equivalent oxygen lattice sites, and hence oxygen vacancies exhibit different
 coordination numbers.$\cite{FosterSGSN01}$ In contrast, all oxygen vacancy sites in
 cubic zirconia are equivalent, and have four-fold coordination.

 In accord with previous work, we find that point defects significantly distort the nearby
 lattice.$\cite{FosterSN02,KralikCL98,FabrisPF02}$ For different charge states the lattice
 relaxations around the vacancy are qualitatively different. Formation of neutral vacancy,
 V$_{\rm o}$ (using Kroger-Vink notation$\cite{KrogerV57}$) results in inward displacement
 of the nearest neighbors. Allowing the volume of supercell to relax results in decrease in
 total volume by 1.03$\rm{\AA}^{3}$. The neighboring Zr and oxygen ions move towards the
 oxygen vacancy. In contrast to V$_{\rm o}$, the atomic structures of V$_{\rm o}^{1+}$ and
 V$_{\rm o}^{2+}$ have outward movement of Zr ions while the oxygen ions move inwards.
 With increase in charge on the vacancy we find increase in distortion of the lattice.
  
 The relaxation volume of a charged vacancy has contributions from the local atomic
 relaxation around the vacant lattice site as well as the delocalized electrons
 present in the system. For a doubly charged oxygen vacancy the net change in volume
 is $\Delta$Vol(V$_{\rm o}^{2+}$) = $\Delta$Vol(ZrO$_{2}$+V$_{\rm o}^{2+}$) +
 2$\Delta$Vol(ZrO$_{2}$+e$^{-}$). Here,
 $\Delta$Vol(ZrO$_{2}$+e$^{-}$) is the change in volume of the ZrO$_{2}$ cell with
 an extra electron and $\Delta$Vol(ZrO$_{2}$+V$_{o}^{2+}$) is the change in volume
 after incorporating a doubly charged vacancy (cell has an effective charge of +2e).
 This calculation yields $\Delta$Vol(V$_{\rm o}^{2+}$) = 5.51$\rm{\AA}^{3}$ and
 $\Delta$Vol(V$_{\rm o}^{1+}$) = 2.17$\rm{\AA}^{3}$. The decrease in energy during this
 relaxation process is 0.09 eV for V$_{o}^{2+}$, 0.32 eV for V$_{o}^{1+}$ and 0.41 eV for
 V$_{o}$ (the cost of simultaneously putting electrons in the conduction band is not
 included). The simulation cell following volume relaxation preserves the cubic
 symmetry. Thus vacancies tend to stabilize the cubic phase in ZrO$_{2}$.

 The process of vacancy formation (say V$_{\rm o}^{2+}$) can be described by the
 following equation :
 \begin{equation}
   \rm O_{o} \Longleftrightarrow \frac{1}{2}O_{2}\left(gas\right) + V_{o}^{2+} + 2e^{-}
 \end{equation}
 Physically, this means the atmosphere acts as sink for the oxygen atoms removed from
 ZrO$_{2}$. The removal of a neutral oxygen atom from the system (leaving behind a
 neutral ZrO$_{2}$ lattice) results in a vacant lattice site. The vacant lattice site
 can trap electrons to form a neutral vacancy or singly charged vacancy. 

 Fig. $\ref{formation0}$ and Fig. $\ref{formation2600}$ shows the formation energy of
 oxygen vacancy as a function of Fermi level for the charge states V$_{\rm o}$,
 V$_{\rm o}^{1+}$, V$_{\rm o}^{2+}$ at $T$ = 0 K and $T$ = 2600 K. When the Fermi level
 is near the valence band edge,
 V$_{\rm o}^{2+}$ is the most stable charge state of a vacancy, but at higher Fermi
 levels neutral charge state (V$_{\rm o}$) becomes energetically favored. We define
 $\epsilon\rm\left(V^{2+}_{o}/V^{1+}_{o}\right)$ as the point at which V$_{\rm o}^{2+}$
 and V$_{\rm o}^{1+}$ have similar values of formation energy. It signifies the
 transition in thermodynamic stability of doubly charged vacancy to a singly
 charged vacancy with increase in Fermi energy. Also from Fig. $\ref{formation0}$ and
 Fig. $\ref{formation2600}$ we can see that the thermodynamic transition
 level$\cite{BlochlS99}$ $\epsilon\rm\left(V_{o}^{2+}/V_{o}^{1+}\right)$ is located
 higher than $\epsilon\rm\left(V_{o}^{1+}/V_{o}^{0}\right)$. Hence, as in the
 monoclinic phase, the formation of V$_{\rm o}^{1+}$ is not favorable in cubic zirconia
 at any Fermi level, meaning that the oxygen vacancy is a negative-$U$
 defect.$\cite{FosterSGSN01}$ The defect levels we find are similar to previous
 estimates using screened
 exchange method and weighted density approximation.$\cite{RobertsonXF05}$

\subsection*{3C. Oxygen Interstitial}
\begin{figure}[thbp]
 \centering
 \subfigure[]  {
    \includegraphics[width=0.3\textheight]{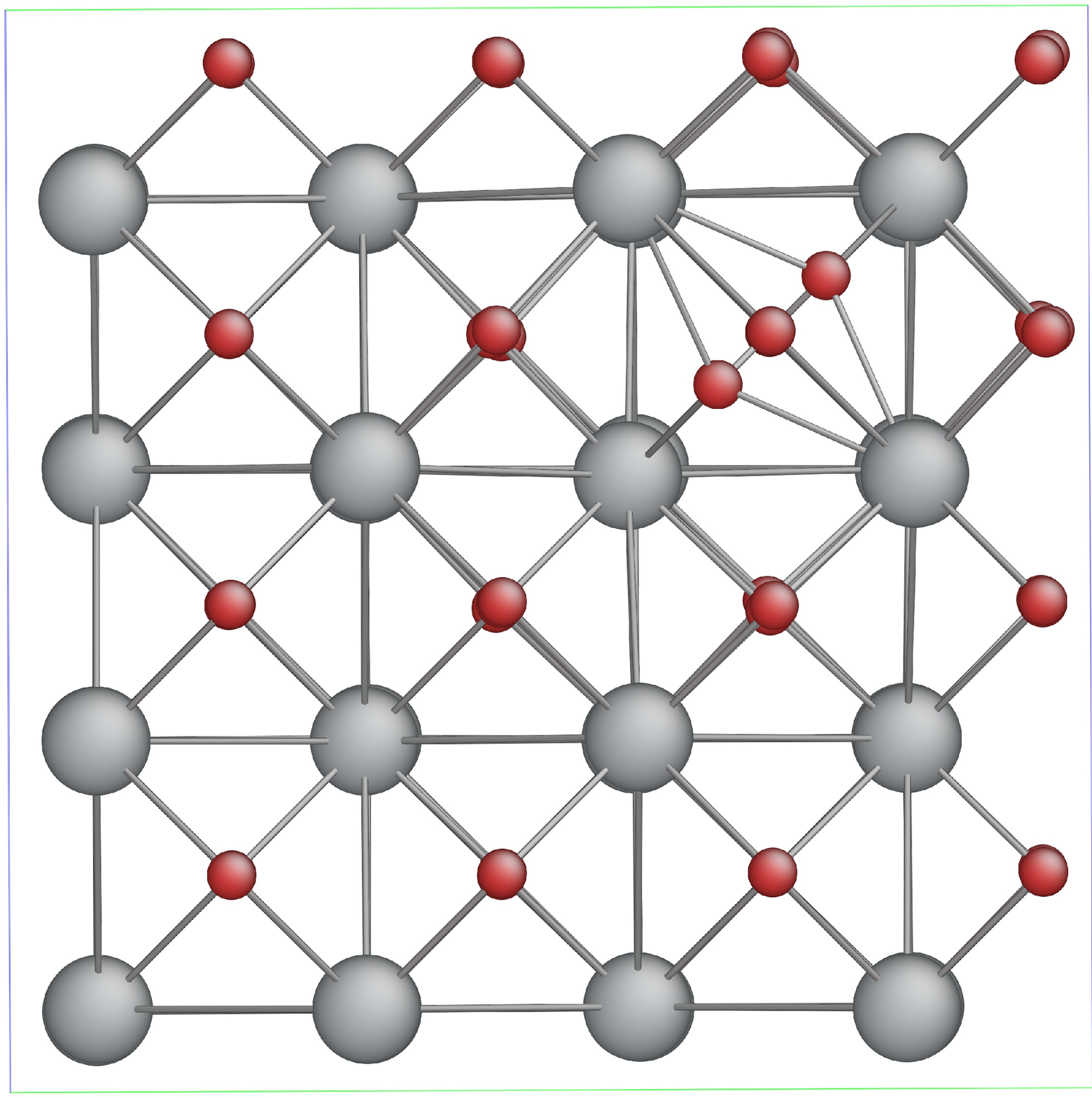}
    \label{110db}
  }
 \subfigure[] {
  \includegraphics[width=0.3\textheight]{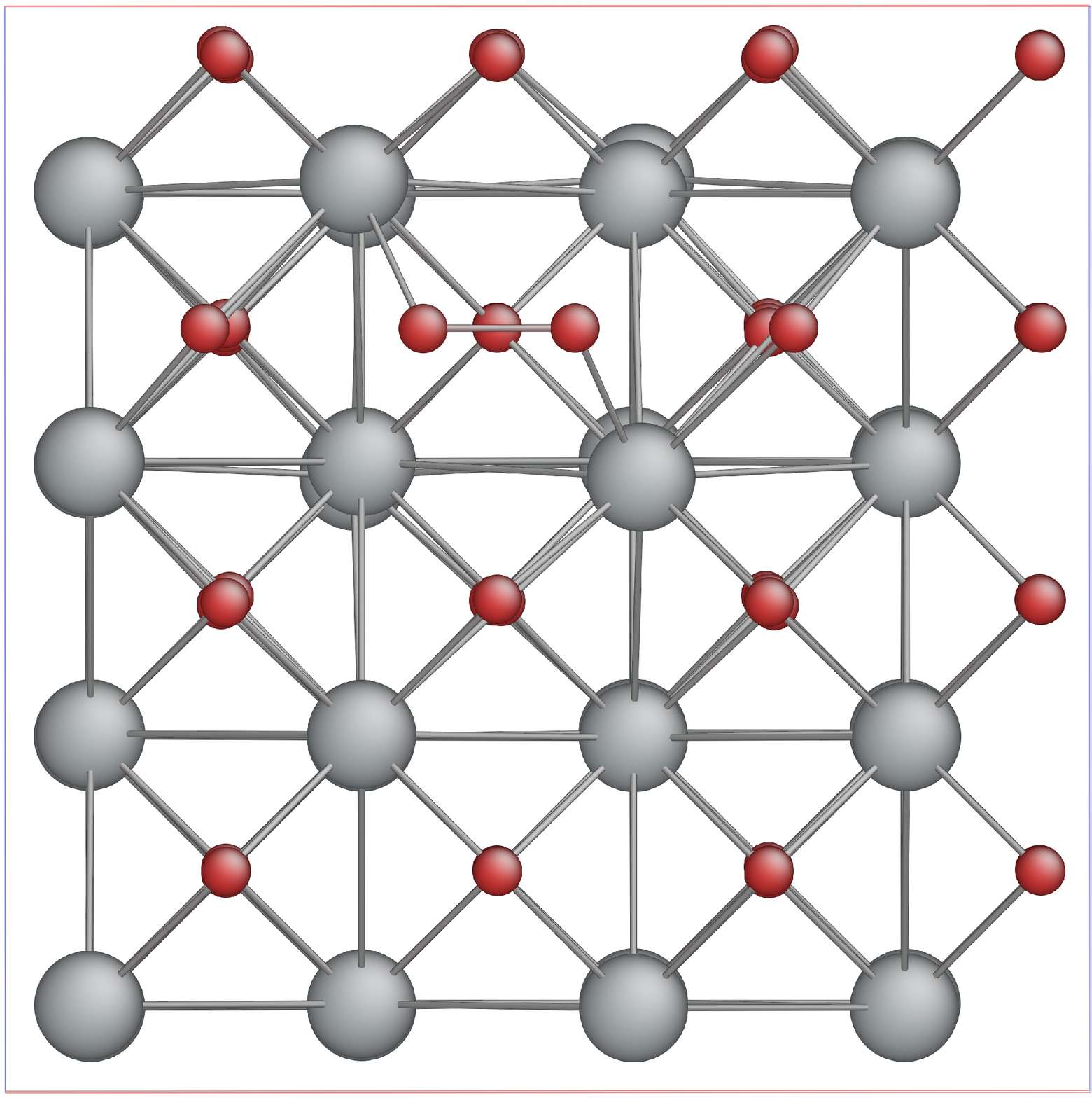}
  \label{100db}
 }
 \vspace{1cm}
 \subfigure[] {
  \includegraphics[width=0.3\textheight]{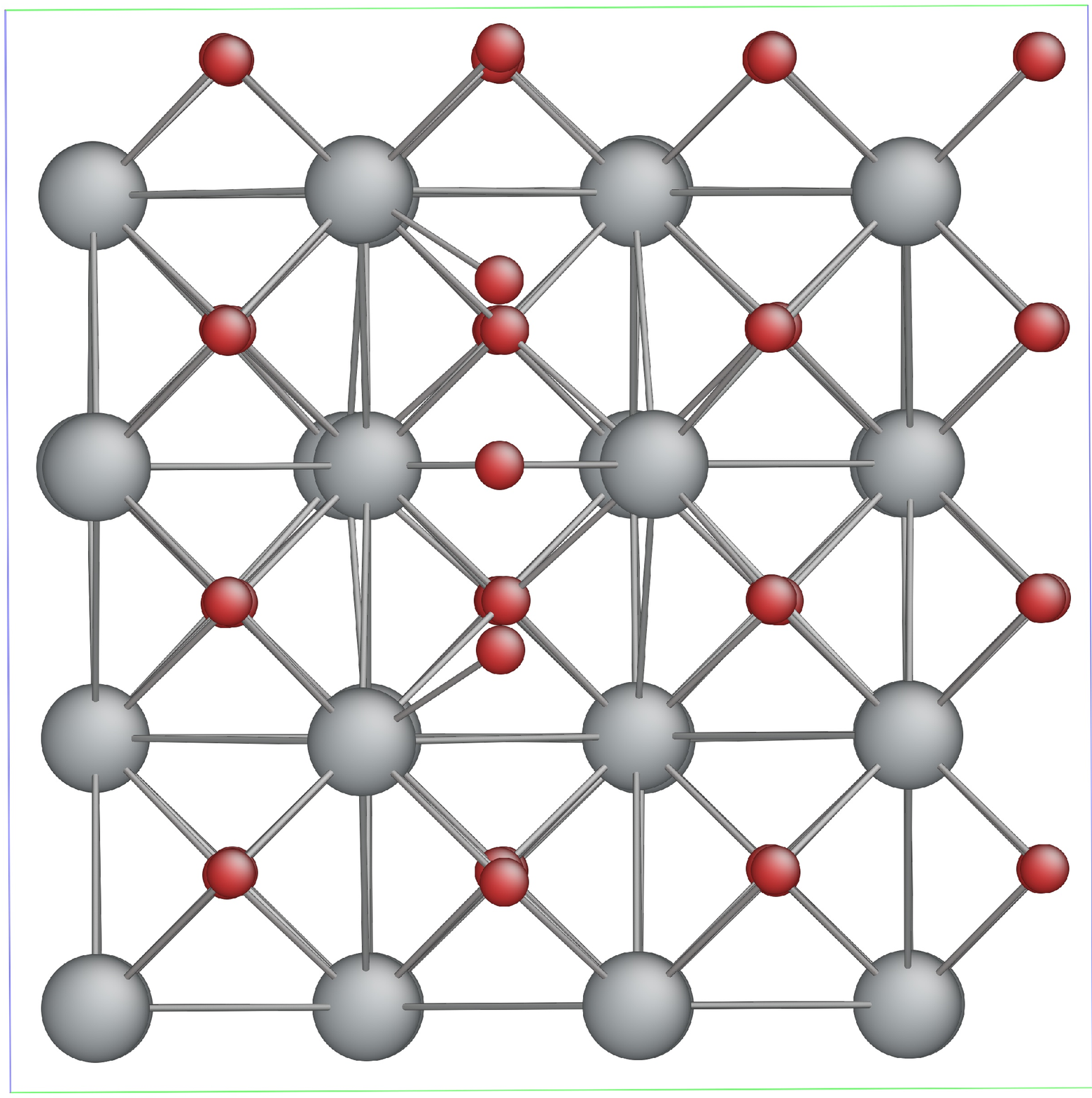}
  \label{100c}
 }
 \subfigure[] {
  \includegraphics[width=0.3\textheight]{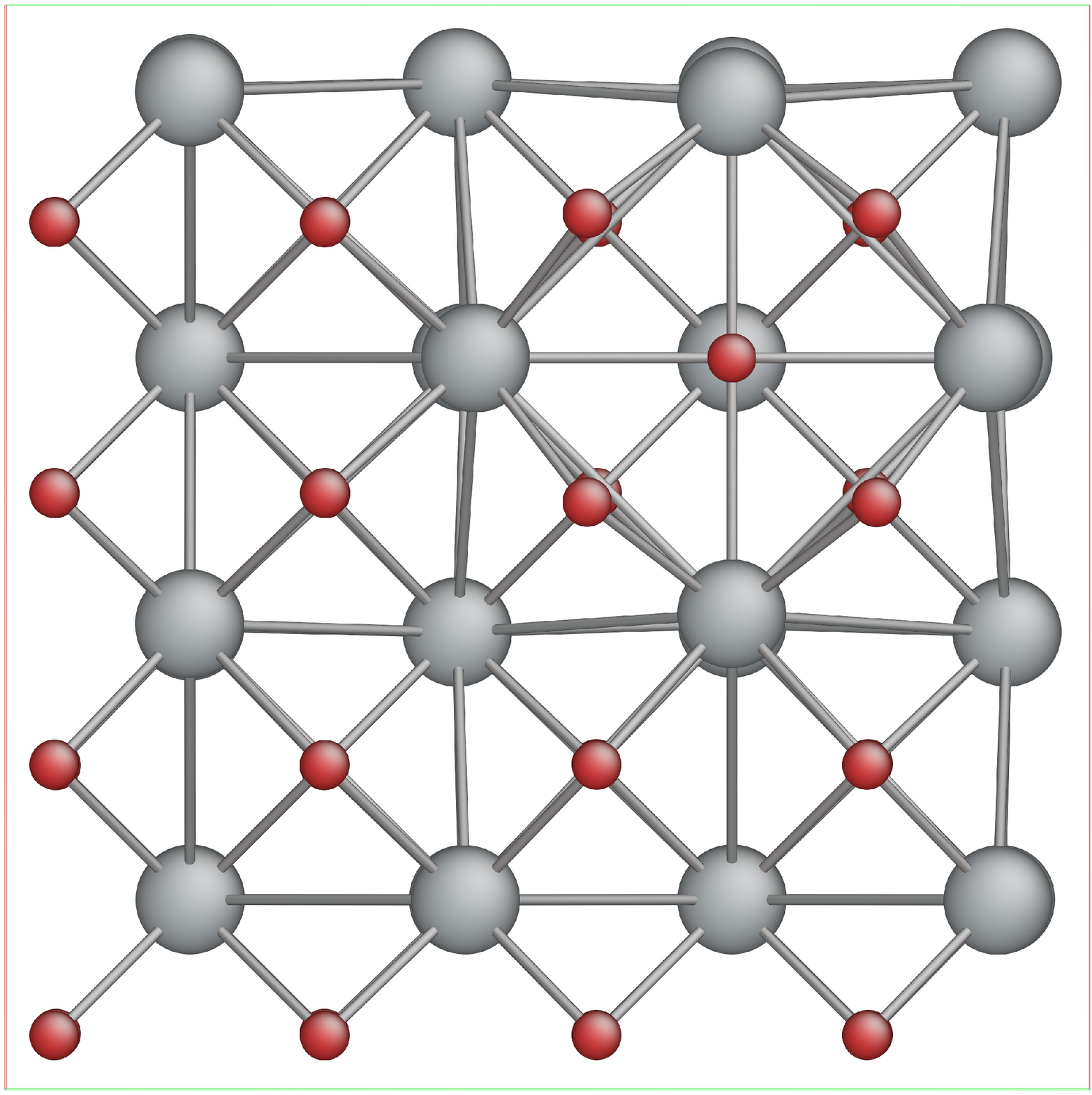}
  \label{octaO2n}
 }
 \caption{Snapshots of oxygen interstitial configurations in cubic zirconia viewed
 along (100) direction. $\ref{110db}$ The $\langle 110\rangle$ dumb bell
 configuration neutral oxygen interstitial. This is the lowest energy configuration
 of the neutral oxygen interstitial, which is energetically favored  at low Fermi
 level. $\ref{100db}$ $\langle 100\rangle$ crowdion configurations of neutral oxygen
 interstitial. $\ref{100c}$ $\langle 100\rangle$ crowdion configurations of neutral
 oxygen interstitial. $\ref{octaO2n}$ Doubly charged oxygen interstitial in octahedral
 configuration. This is the ground state of a doubly charged oxygen interstitial.}
 \label{Config}
 \end{figure}

 Cubic ZrO$_{2}$ has CaF$_{2}$ structure with metal atoms in a face
 centered cubic sub-lattice and oxygen atoms occupying a
 simple cubic sub-lattice. Since the oxygen sub-lattice is not
 close packed, oxygen defects, rather than zirconium defects,
 play a dominant role in determining the properties of this
 oxide. We find that oxygen interstitials in cubic zirconia have different
 possible atomic structures than those previous reported for the monoclinic phase. We
 have studied different oxygen interstitial configurations
 and geometries : $(i)$ $\langle 110\rangle$ dumbbell
 (Fig. $\ref{110db}$), $(ii)$ $\langle 100\rangle$ dumbbell
 (Fig. $\ref{100db}$), $(iii)$ $\langle 100\rangle$ crowd-ion
 (Fig. $\ref{100c}$) and $(iv)$ octahedral site (Fig.
$\ref{octaO2n}$). For ease of reference the interstitial
 configurations are labeled as O$_{\langle 110\rangle}$,
 O$_{\langle 100\rangle}$ respectively for the $\langle 110\rangle$,
 and $\langle 100\rangle$ dumbbell configurations,
 O$_{\langle 100\rm c\rangle}$ for $\langle 100\rangle$ crowd-ion and
 O$_{\rm oct}$ for octahedral site. Another possible interstitial configuration
 is $\langle 111\rangle$ dumbbell. While $\langle 110\rangle$ dumbbell
and $\langle 100\rangle$ dumbbell have dumbbell centers on an original atomic site, 
a $\langle 111\rangle$ dumbbell configuration
 upon relaxation has dumbbell center shifted and not on an original atomic site. 
$\langle 111\rangle$ dumbbell in fact 
looks quite
similar to O$_{\rm oct}$, but not equivalent. It has 0.88 eV higher energy than
 O$_{\langle 110\rangle}$ and the dumbbell separation is 1.43 $\rm{\AA}^{3}$.
 
 The relative stability of different interstitial configurations with respect to
 O$_{\langle 110\rangle}$ is compared in Table \ref{structure}. For an oxygen
 interstitial with neutral charge, O$_{\langle 110\rangle}$
 configuration is energetically most favorable. The dumbbell separation is close to
 the O-O bond length in peroxides.$\cite{Tian04, AvdeevRZ97}$ Incorporation of an
 extra oxygen atom results in outward movement of the nearest neighbor
 Zr ions. In this case the lattice distortion is symmetrical about
 $\left( 110\right)$ plane. The Zr ions along $\langle 110\rangle$
 moves 6$\%$ (of the Zr-O bond length in an ideal crystal) outward. The
 nearest neighbor oxygen ions move slightly inward by 0.7$\%$.  The
 two oxygen ions forming the dumbbell structure are separated by
 1.46 $\rm{\AA}$. The change in volume of supercell for
 O$_{\langle 110\rangle}$ was calculated by allowing the supercell
 to relax. For O$_{\langle 110\rangle}$ structure, we find it to be 36.10
 $\rm{\AA}^{3}$, much higher than the values obtained for vacancies. Volume
 relaxation destroys the cubic symmetry of the simulation cell.

 The electronic local density of states plots near one of the dumbbell ions are the same
 for spin up and spin down. This is also true for any oxygen atom in the bulk far away
 from the interstitial configuration. Thus the oxygen ions in the system do not have
 residual spins. Robertson et al. performed
 calculations for oxygen interstitials in cubic zirconia and reported electronic
 properties of O$_{\langle 100\rangle}$, though our findings suggest the
 O$_{\langle 110\rangle}$ as the most stable.

 In order to verify the lowest energy configuration we perform ab-initio molecular
 dynamics (MD) simulations at $T$ = 2000 K for 1.5 ps. The high temperature enhances
 mobility of the ions and the structures obtained are much different from those at
 $T$ = 0 K. Next, we select a few intermediate configurations from the MD simulation
 results and cool them to $T$ = 0 K in molecular dynamics and perform static energy
 minimization. This procedure typically yields significantly lower-energy structures
 ($\Delta E\sim$ 0.5 eV) for vacancies and interstitials. These low energy structures
 have low symmetry (due to Jahn-Teller distortion). This methodology is well-behaved
 in that when the same procedure is applied to a perfect cubic zirconia cell, the
 perfect cell is recovered. When performing this procedure on defects we do not allow
 the cell lattice vectors to relax, since if this is allowed the cell can become
 tetragonal.  

 \begin{table}[tbh]
 \begin{tabular}{c|c|c|c|c|c|c}
 \hline
 \hline
 Defect $\;$& $\;$ Interstitial $\;$& $\;$ O$_{\langle 110\rangle}$ $\;$& $\;$
 O$_{\langle 111\rangle}$ $\;$& $\;$ O$_{\langle 100\rangle}$ $\;$& $\;$
 O$_{\langle 100\rm c\rangle}$ $\;$ &$\;$ O$_{\rm oct}$ \\
 \hline
 O$^{0}_{\rm i}$ & Energy (relative, eV) & 0 & 0.88 & 0.95 & 1.92 & 3.02 \\
  & Dumbbell separation ($\rm{\AA}$) & 1.46 & 1.44 & 1.43 & - & - \\
 \hline
 O$^{1-}_{\rm i}$ & Energy (relative, eV) & 0 & - & 3.80 & 3.79 & 1.90 \\
  & Dumbbell separation ($\rm{\AA}$) & 1.98 & - & 1.93 & - & - \\
 \hline
 \hline
 \end{tabular}
 \caption{Relative stability and dumbbell
 separations for neutral and singly charged oxygen interstitials in
 cubic zirconia. After annealing by ab-initio MD and possessing lower symmetry due to Jahn-Teller distortion.}
 \label{structure}
 \end{table}

 For singly charged interstitials, we find that the O$_{\langle 110\rangle}^{1-}$
 configuration is energetically most favorable in contrast to O$_{\rm oct}^{1-}$
 reported previously by Robertson et al.$\cite{RobertsonXF05}$ We come to this
 conclusion by performing MD simulations at $T$ = 2000 K and then quenching to $T$ =
 0 K. After static relaxation, we find the separation between the ions in a dumbbell
 is 1.98 $\rm{\AA}$, much higher than that in a neutral
 cell. The O$_{\langle 111\rangle}^{1-}$ configuration becomes unstable and the
 interstitial atom moves to nearest octahedral site. Table \ref{structure} shows
 the relative stability of different configurations for a singly charged interstitial
 with respect to O$_{\langle 110\rangle}^{1-}$.

 The most stable site for O$_{\rm i}^{2-}$ is the octahedral site. Interstitial
 atoms placed at the other configurations become unstable and move to occupy the
 nearest octahedral site upon relaxation. Similar to the neutral oxygen interstitial,
 the projected density of states for spin-up and spin-down are identical and hence
 there is no residual spin on the interstitial atom. Similar to the case of charged
 vacancies, incorporating charged interstitials, O$_{\rm i}^{2-}$ results in large
 distortion of the parent lattice. The neighboring oxygen and zirconium ions display
 outward distortions. 

 Physically, the process of formation of O$_{\rm i}^{2-}$ can be understood by this
 reaction :
 \begin{equation}
   \rm \frac{1}{2}O_{2}\left(gas\right)\Longleftrightarrow O_{i}^{2-} + 2h^{1+}.
 \end{equation}
 Here, the atmosphere acts as source of oxygen. The system has to be neutral before and
 after the formation of the interstitial, so formation of a charged interstitial results
 in the formation of holes. Fig. $\ref{formation0}$ and
 Fig. $\ref{formation2600}$ shows the formation energy of O$_{\rm i}^{2-}$,
 O$_{\rm i}^{1-}$ and O$_{\rm i}^{0}$ as a
 function of the Fermi energy of the system at 1 atm pressure and $T$ = 0 K and $T$
 = 2600 K respectively. With increasing temperature incorporation of oxygen
 interstitials become energetically expensive. For example, the incorporation of
 neutral interstitial at $T$ = 0 K costs $E$ = 0.86 eV but at $T$ = 2600 K is $E$
 = 4.12 eV. From the formation energy plot, we conclude that the thermodynamic defect
 level $\epsilon\rm\left(O_{i}^{0}/O_{i}^{1-}\right)$ at $T$ = 0 K is higher
 than $\epsilon\rm\left(O_{i}^{1-}/O_{i}^{2-}\right)$. Hence, as in the monoclinic
 phase$\cite{FosterSN02}$, O$_{\rm i}^{1-}$ is energetically not favored at any
 Fermi level in cubic ZrO$_{2}$, meaning that the oxygen interstitial is a
 negative-$U$ defect.

 \begin{table}[tbh]
 \begin{tabular}{c|c|c} 
 \hline
 \hline
 No. $\;$& $\;$ Reaction $\;$& $\;$ Energy (eV) \\
 \hline
 1 & O$_{\rm i}^{2-}$+O$_{\rm i}^{0}$$\Rightarrow$2O$_{\rm i}^{1-}$ & 0.83 \\
 2 & V$_{\rm o}^{2+}$+V$_{\rm o}^{0}$$\Rightarrow$2V$_{\rm o}^{1+}$ & 0.43 \\
 3 & Null $\Rightarrow$O$_{\rm i}^{0}$+V$_{\rm o}^{0}$ & $E_{\rm F}^{''}$=5.15 \\
 4 & Null $\Rightarrow$O$_{\rm i}^{1-}$+V$_{\rm o}^{1+}$ & $E_{\rm F}^{'}$=4.08 \\
 5 & Null $\Rightarrow$O$_{\rm i}^{2-}$+V$_{\rm o}^{2+}$ & $E_{\rm F}$=1.75 \\
 6 & O$_{\rm i}^{0}$+V$_{\rm o}^{0}$$\Rightarrow$O$_{\rm i}^{1-}$+V$_{\rm o}^{1+}$ & -1.07 \\
 7 & O$_{\rm i}^{1-}$+V$_{\rm o}^{1+}$ $\Rightarrow$O$_{\rm i}^{2-}$+V$_{\rm o}^{2+}$ & -2.33 \\
 8 & O$_{\rm i}^{0}$+V$_{\rm o}^{0}$ $\Rightarrow$O$_{\rm i}^{2-}$+V$_{\rm o}^{2+}$ & -3.40 \\
 9 & Null $\Rightarrow$e$^{-}$+h$^{+}$ & $E_{\rm g}$=5.4 \\
 \hline
 \hline
 \end{tabular}
\caption{Defect reactions and the corresponding energy change at $T$ = 0 K.
 The energy change is obtained from the formation energy of the defects. The
 values reported have been corrected for dipole interactions.}
\label{DefectReactionsVASP1}
 \end{table}

 Table $\rm{\ref{DefectReactionsVASP1}}$ shows some of the possible reactions between
 the anionic defects in ZrO$_{2}$ and their reaction energies. These reaction energies
 are calculated at 0 K and zero pressure assuming dilute concentration of the defect
 species. A positive energy means the forward reaction is unfavorable. Reactions 1 and
 2 have positive reaction energies, which means that the formation of singly charged
 interstitials and vacancies are not favored. Thus, two singly charged interstitials
 or vacancies
 would decay into neutral and doubly charged defects. Reactions 6-8 suggest that a
 pair of neutral vacancy and interstitial would prefer to form a pair of doubly
 charged Frenkel defect.

 Next we study the incorporation of an oxygen molecule. The formation energy of an
 oxygen molecule in cubic ZrO$_{2}$ is 10.42 eV at 1 atm oxygen partial pressure
 and $T$ = 2600 K. At zero pressure and $T$ = 0 K the formation energy is 4.51 eV,
 much higher
 than interstitial formation energy of 0.86 eV in a neutral cell. The oxygen molecule
 formation energy in cubic ZrO$_{2}$ is close to the values reported for low
 temperature monoclinic phase of HfO$_{2}$ and much higher than
 those reported to silica.$\cite{FosterGSN02}$ The bond length of the
 oxygen molecule is 1.27 $\rm\AA$. By performing MD simulations at 2600K
 for 1ps and then subsequent relaxation, we find that the molecule
 prefers to dissociate into two $\langle 110\rangle$ dumbbells. The
 corresponding decrease in energy obtained from static relaxation is 1.93 eV.

\section*{4. Defect Equilibrium Reactions}
 The thermodynamic stability of different point defects can be understood by writing
 down the equilibrium chemical reactions for the formation of these defects. Analysis
 of the defect chemistry is done by obtaining the laws of mass
 action for different competing processes.$\cite{Huggins01, BogdanovDLT98, Barsoum97}$
 Apart from the laws of mass action, the various point defects present must also satisfy
 the electronic charge neutrality condition. Here we consider the stable anionic defects
 in ZrO$_{2}$, namely O$_{\rm i}$, O$_{\rm i}^{2-}$, V$_{\rm o}$ and V$_{\rm o}^{2+}$.
 Each defect is considered to be in its lowest energy configuration. In cubic ZrO$_{2}$,
 the metal ions occupy the closed packed face centered lattice sites. So cationic
 defect formation energies are much higher than anionic defect formation
 energies.$\cite{ZhengCMCC07}$ Hence we neglect the cationic defect contributions in
 subsequent analysis.

 Possible reactions involving O$_{\rm i}^{2-}$ and V$_{\rm o}^{2+}$ are as follows:\\           
$\bullet$ formation of Frankel defects in the anionic sub-lattice : 
\begin{equation}
  {\rm O_{\rm o} \Longleftrightarrow O_{\rm i}^{2-} + V_{\rm o}^{2+}}
  \qquad\qquad\qquad\qquad\qquad K_{\rm F} = 
  \frac{\left[\rm{O_{i}}^{2-}\right]\left[\rm{V_{o}}^{2+}\right]}{N_{\rm a}^{2}}
\label{FrankelO2n}
\end{equation}
 $\bullet$ formation of an oxygen vacancy by losing oxygen to the
 atmosphere : 
\begin{equation}
  {\rm O_{o} \Longleftrightarrow \frac{1}{2}O_{2}\left(gas\right) +
    V_{o}^{2+} + 2e^{-}} \qquad\qquad\qquad K_{1} =
  \frac{\left[{\rm V_{o}^{2+}}\right]\left[{\rm e^{-}}\right]^{2}}{N_{\rm a}N_{\rm c}^{2}}
  P_{\rm{O}_{2}}^{1/2}
  \label{O2nRed}
\end{equation}
 $\bullet$ formation of electron-hole pair :
\begin{equation}
  {\rm Null \Longleftrightarrow e^{-} + h^{+}} \qquad\qquad\qquad\qquad K_{\rm i} =
  \frac{\left[{\rm h^{+}}\right]\left[{\rm e^{-}}\right]}{N_{\rm v}N_{\rm c}} = 
  e^{-E_{\rm g}/k_{\rm B}T}
\end{equation}
$\bullet$ formation of oxygen interstitial :
\begin{equation}
  {\rm \frac{1}{2}O_{2}\left(gas\right) \Longleftrightarrow O_{i}^{2-} +
    2h^{+}} \qquad\qquad\qquad K_{2} =
  \frac{\left[{\rm h^{+}}\right]^{2}\left[{\rm O_{i}^{2-}}\right]}
       {N_{\rm v}^{2}N_{\rm a}P_{\rm O_{2}}^{1/2}}
       \label{O2nOxid}
\end{equation}
 $\bullet$ and charge neutrality condition :
\begin{equation}
  \rm 2\left[O_{i}\right]^{2-} + \left[e^{-}\right] = 2\left[V_{o}\right]^{2+} +
  \left[h^{+}\right]
  \label{chgNeutralO2n}
\end{equation}
 where, $\left[\rm e^{-}\right]$ and $\left[\rm h^{+}\right]$ are the electron and hole
 concentration,
 respectively, $\left[{\rm O_{i}}^{2-}\right]$ is the oxygen interstitial concentration,
 $\left[{\rm V_{o}}^{2+}\right]$ is the oxygen vacancy concentration, $N_{\rm a}$ is the
 concentration of anionic sites and $E_{\rm g}$ is the band gap. $N_{\rm c} =
 2\left(2\pi m_{\rm e}^{*}k_{\rm B}T/h^{2}\right)^{3/2}$ and $N_{\rm v} =
 2\left(2\pi m_{\rm h}^{*}k_{\rm B}T/h^{2}\right)^{3/2}$ are the constants used to
 normalize the electron and hole concentration.$\cite{Barsoum97}$

 Since $K_{\rm F}=e^{-E_{\rm F}/k_{\rm B}T}$ and $K_{\rm i}=e^{-E_{\rm g}/k_{\rm B}T}$, using
 values from Table \ref{DefectReactionsVASP1}, we find $K_{\rm F}\gg$ $K_{\rm i}$. Hence
 the concentration of electronic defects $\left[\rm h^{+}\right]$, $\left[\rm e^{-}\right]$
 are much smaller than the doubly charged Frankel defects. In this limit, at intermediate
 oxygen partial pressure, the charge neutrality equation can be simplified
 (Brouwer approximation$\cite{Brouwer54}$) to
 $\left[\rm {O_{i}}^{2-}\right]=\left[{\rm V_{o}}^{2+}\right]\propto\sqrt{K_{\rm F}}$.
 When the oxygen partial pressure is very high, $\left(\ref{O2nOxid}\right)$ shifts
 towards right and we get more oxygen interstitials. So using the charge neutrality
 $\left(\ref{chgNeutralO2n}\right)$ condition, the concentration of oxygen interstitials
 can be expressed in terms of the oxygen partial pressure:

 \begin{equation}
   \left[{\rm V_{\rm o}}^{2+}\right] \propto P_{\rm O_{2}}^{-1/6} \qquad\qquad\qquad\qquad
   \left[{\rm O_{i}}^{2-}\right] \propto P_{\rm O_{2}}^{1/6}.
 \end{equation}
 
 The defect concentrations are dependent on multiple factors. In nature, the defect
 concentrations can be changed by adding dopant or by varying the external environment
 (temperature and pressure). Here we take undoped zirconia and assume that the defect
 concentrations, are determined by these two external degrees of freedom.

 For the neutral defects the thermodynamic equilibrium reactions for
 O$_{\rm i}$ and V$_{\rm o}$ are given by,\\
 $\bullet$ formation of Frankel defects in the anionic sub-lattice :
 \begin{equation}
 {\rm O_{o} \Longleftrightarrow O_{i} + V_{o}} \qquad\qquad\qquad K_{\rm F}^{''} =
 \frac{{\rm \left[O_{i}\right]\left[V_{o}\right]}}{N_{\rm a}^{2}}
 \end{equation}
 $\bullet$ formation of an oxygen vacancy by loosing oxygen to the
 atmosphere :
 \begin{equation}
   {\rm O_{o} \Longleftrightarrow \frac{1}{2}O_{2}\left(gas\right) + V_{o}}
 \qquad\qquad\qquad K_{1}^{''} =
 \frac{\left[{\rm V_{o}}\right]P_{\rm O_{2}}^{1/2}}{N_{\rm a}}
 \end{equation}
 $\bullet$ formation of oxygen interstitial :
 \begin{equation}
   {\rm \frac{1}{2}O_{2}\left(gas\right) \Longleftrightarrow O_{i}}
   \qquad\qquad\qquad K_{2}^{''} =
   \frac{\left[{\rm O_{i}}\right]}{N_{\rm a}P_{\rm O_{2}}^{1/2}}
 \end{equation}
 Analyzing the system for neutral defects is much simpler than the charged defects.
 Solving the above we get
 \begin{equation}
 {\rm \left[O_{i}\right]}\propto P_{O_{2}}^{1/2} \qquad\qquad\qquad\qquad
 {\rm \left[V_{o}\right]}\propto P_{O_{2}}^{-1/2}
 \end{equation}

\section*{5. Determination of equilibrium Fermi energy}
 The formation energy plots provide information about the stability of different point defects as a 
function of temperature, pressure and Fermi energy of the system. Consider for example, defect 
formation energies at 2600 K and 1 atm (Fig. $\ref{formation2600}$). The doubly charged defects have 
the lowest energies. However, for Fermi energy values less than $\sim$2.1 eV and more than 3 eV the 
system becomes unstable leading to spontaneous formation of defects. This pins the Fermi energy of 
the system to a small region of 2.1 to 3 eV. The overall system needs to maintain charge neutrality. 
For this to happen the Fermi energy ($\mu_{\rm e}$) of the system needs to be 2.6 eV so that doubly 
charged anionic defects have same concentration.

In order to understand this better let us look at the charge neutrality condition for undoped ZrO$_{2}$ 
described in ($\ref{chgNeutralO2n}$). We neglect the singly charged defects as they are thermodynamically 
not stable. We do not include the cationic defects as higher formation energy would make their concentration 
insignificant. As discussed in before, we can have three regimes : (a) $\left[{\rm e^{-}}\right] = 
2\left[{\rm V_{o}}\right]^{2+}$ - this implies significant electronic conductivity which is not physical 
for an insulator like ZrO$_{2}$; (b) $\rm \left[h^{+}\right] = 2\left[O_{i}\right]^{2-}$ - similar to the 
above condition this is not possible; (c) $\rm \left[O_{i}\right]^{2-} = \left[V_{o}\right]^{2+}$ - this 
regime seems physically plausible.
   
\begin{figure}[thbp]
   \centering
   \includegraphics[width=0.5\textheight]{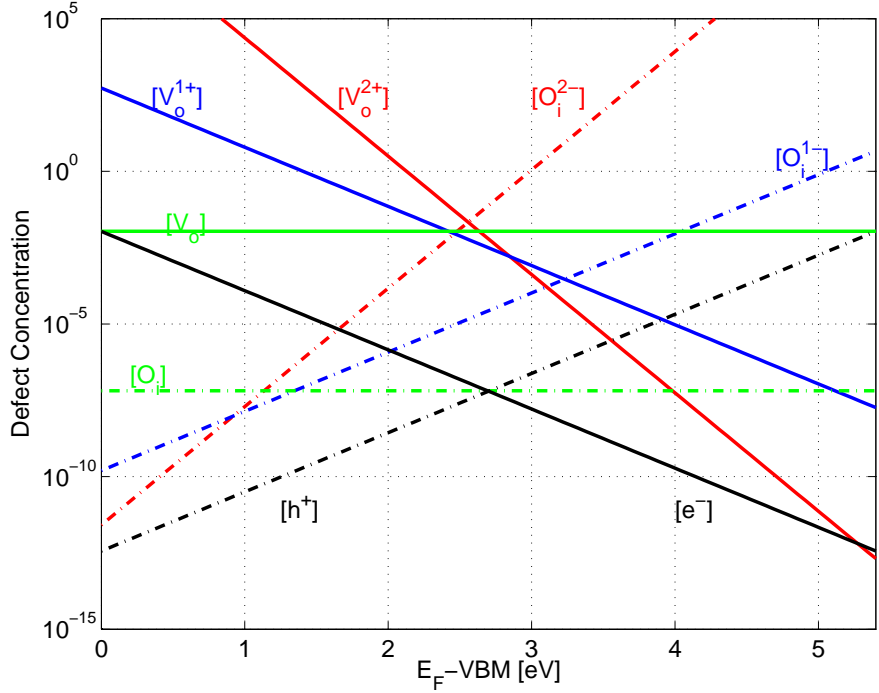}
 \caption{Comparison of point defect concentration as a function of Fermi level of the
   system at 1 atm pressure and $\ref{ConcFermiLevel2600K}$ $T$ = 2600 K. The
   concentration profiles are obtained by solving each set of defect equilibrium
   equations separately. The doubly charged
   defects : V$_{\rm o}^{2+}$ (solid red line) and O$_{\rm i}^{2-}$ (dotted red line)
   have higher concentration than the singly charged (V$_{\rm o}^{1+}$ - solid blue
   line, and O$_{\rm i}^{1-}$ - dotted blue line)and neutral defects (V$_{\rm o}$)
   - slid green line, and O$_{\rm i}$ - dotted green line. The defect concentrations
   are expressed as fraction of anionic sites in the system. }
   \label{ConcFermiLevel2600K}
 \end{figure}

In order to confirm this conclusion, we plot the defect concentrations at 2600 K and 1 atm pressure as 
a function of the Fermi energy. Fig. $\ref{ConcFermiLevel2600K}$ shows a comparison between the defect 
concentrations and the electron and hole concentrations. The defect concentrations are calculated from 
the rate constants derived from the laws of mass action, namely, ($\ref{O2nRed}$) and ($\ref{O2nOxid}$). 
The electron and hole concentrations are obtained by $\left[{\rm e^{-}}\right] = N_{\rm c}e^{-\left(E_{\rm c}- 
\mu\right)/k_{\rm B}T}$ and $\left[{\rm h^{+}}\right] = N_{\rm v}e^{-\left(\mu - E_{\rm c}\right)/k_{\rm B}T}$. 
Clearly, the doubly charged defects have much higher concentration than the neutral and singly charged 
defects. As discussed above, this analysis shows that the electron and hole concentrations are much 
smaller than the anionic defect concentrations and hence conductivity is by migration of ions.

\section*{6. Conclusions}

An ab-initio study of anionic point defects in cubic zirconia are reported. For O$_{\rm i}$ and 
O$_{\rm i}^{1-}$ the energetically favorable configuration is $\langle 110\rangle$ dumbbell and 
for O$_{\rm i}^{2-}$ it is an octahedral site. Using a combination of high temperature molecular 
dynamics and static minimization we obtain the global minima of the potential energy for these 
defects. Our ground state configurations differ substantially from those previously reported by 
Robertson et al.$\cite{RobertsonXF05}$

Oxygen interstitials and vacancies in the cubic phase exhibit negative-$U$ behavior, meaning that 
the singly charged defects are suppressed at any Fermi level. From the analysis of possible 
reactions between these anionic defects we find that singly charged defect species are thermodynamically 
less preferable than a combination of neutral and doubly charged defects. Using molecular O$_{2}$ as 
the reference state, defect formation energies are obtained at 0 K and 2600 K. The neutral interstitials 
have $\sim$ 2 eV higher formation energy than the vacancies at 2600K and 1 atm pressure. The system can 
lower its energy if the oxygen ions forming the interstitial dumbbell diffuse to the surface and move 
out to the atmosphere as neutral oxygen molecule. This results in the formation of a neutral vacancy. 
The corresponding change in energy is $\Delta$E = E(ZrO$_{2}$+V$_{\rm o}$) - [E(ZrO$_{2}$+O$_{\rm i}$) 
- E(O$_{2}$)] = -2.49 eV. Hence, we conclude that neutral interstitials are not stable in cubic zirconia.

We find the relaxation volumes for the anionic vacancies are much smaller than the oxygen interstitials. 
Unlike the oxygen interstitials, the local atomic relaxation around oxygen vacancies tend to stabilize 
the cubic phase. We find a charged vacancy has considerably higher relaxation volume compared to a neutral 
vacancy. Higher relaxation volume signifies that ZrO$_{2}$ is susceptible to diffusional creep, which can 
degrade its mechanical properties with increase in temperature.

Zirconium based materials are attractive for nuclear waste disposal. The Zr-O binary phase 
diagram$\cite{AckermannGR77}$ shows the existence of non-stoichiometric ZrO$_{2-\delta}$ phase 
suggesting vacancies are dominant defect species. Assuming vacancies are dominant defects, we 
find that change in volume ($\Delta V$) due to presence of defects in cubic ZrO$_{2}$ is not 
large ($\Delta$V/V$<$0.5$\%$). Similar results have also been obtained for ZrSiO$_{4}$.$\cite{PrunedaAA04}$

\begin{figure}[thbp]
 \centering
 \subfigure[]{\includegraphics[width=0.45\textheight]{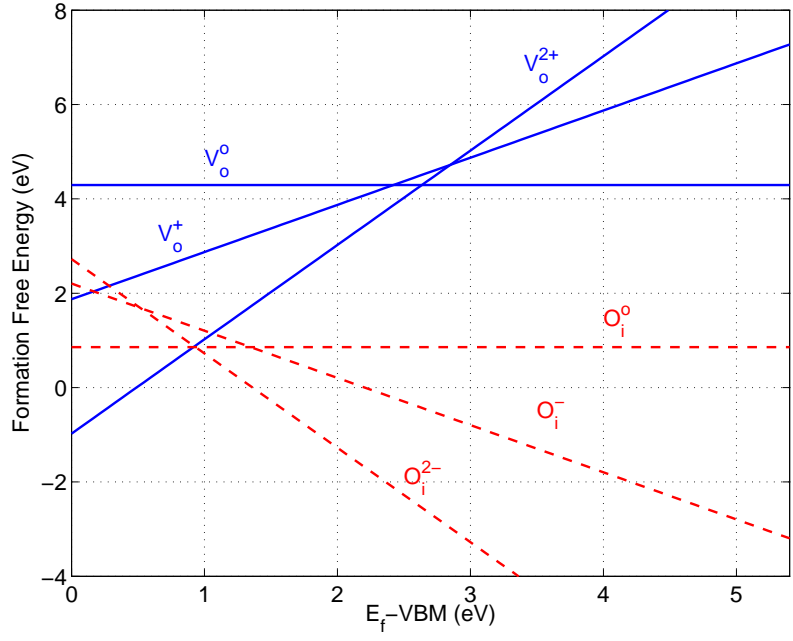}
    \label{formation0}}
 \subfigure[]{\includegraphics[width=0.45\textheight]{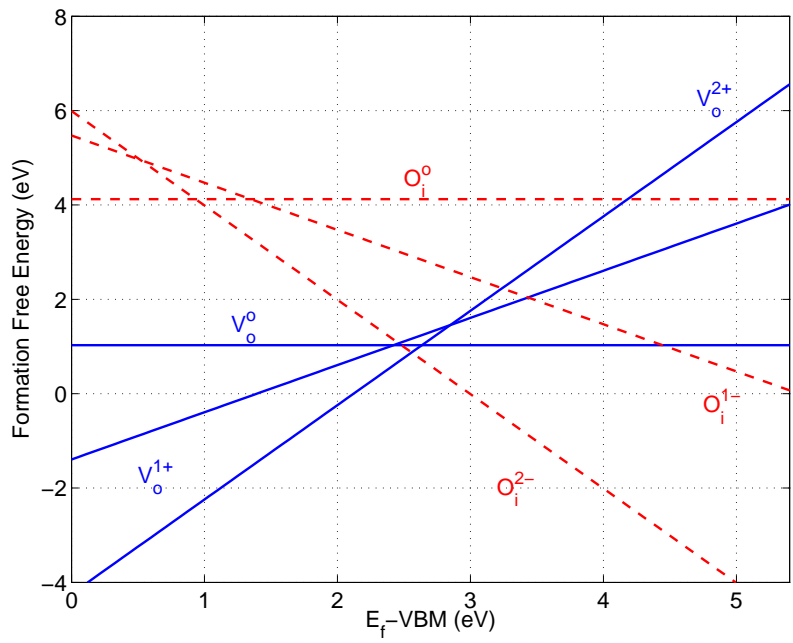}
  \label{formation2600}} 
 \caption{Formation energy of point defects in cubic ZrO$_{2}$ as function of the Fermi
 energy at $\ref{formation0}$ $T$ = 0 K and $\ref{formation2600}$ $T$ = 2600 K and 1 atm.}
 \label{formation}
 \end{figure}

For a stable system, formation energy of a defect has to be positive. This puts a constrain on the 
accessible Fermi energy values. From Fig. $\ref{formation}$ we see that at 0 K, the feasible Fermi 
energy values are $\sim$0.5-1.5 eV, while at 2600 K Fermi energy of the system lies in the range of 
$\sim$2.1-3.0 eV. For undoped ZrO$_{2}$, the Fermi energy would lie mid-gap, however, the Fermi 
energy values at low temperature suggest that concentration of doubly charged defects can be tuned 
by doping with acceptor impurities.

Equilibrium defect concentrations in undoped ZrO$_{2}$ are obtained as a function of oxygen partial 
pressure and temperature by solving the laws of mass action. Based on the charge neutrality condition a 
method to calculate the Fermi energy of the system is proposed. It is important to note that these 
defect concentrations are evaluated without accounting for defect interactions. Thus, in the high 
concentration limit, the actual concentrations can be different from those shown in the figure.

Our results are also relevant to cubic stabilized zirconia (CBZ), in which a p-type dopant such as Y 
or Ca lower the Fermi level. These low values of Fermi level signifies the abundance of doubly charged 
oxygen vacancies. Thus, our results demonstrate that, by solving defect equilibrium equation separately 
and then comparing their concentrations, an estimate of the dominant defect species can be made.

\begin{acknowledgments}
We acknowledge support by the Ohio Supercomputer Center and AFOSR. T. Lenosky wishes to
 thank Paul Erhart for sharing with him a preprint on point defects in BaTiO$_{3}$
 prior to publication. A. Samanta wishes to thank Ji Feng for his comments.
\end{acknowledgments}

\bibliography{MyBibliography}

\begin{thebibliography}{40}
\expandafter\ifx\csname natexlab\endcsname\relax\def\natexlab#1{#1}\fi
\expandafter\ifx\csname bibnamefont\endcsname\relax
  \def\bibnamefont#1{#1}\fi
\expandafter\ifx\csname bibfnamefont\endcsname\relax
  \def\bibfnamefont#1{#1}\fi
\expandafter\ifx\csname citenamefont\endcsname\relax
  \def\citenamefont#1{#1}\fi
\expandafter\ifx\csname url\endcsname\relax
  \def\url#1{\texttt{#1}}\fi
\expandafter\ifx\csname urlprefix\endcsname\relax\def\urlprefix{URL }\fi
\providecommand{\bibinfo}[2]{#2}
\providecommand{\eprint}[2][]{\url{#2}}

\bibitem[{\citenamefont{Zhao and Vanderbilt}(2002)}]{ZhaoV02}
\bibinfo{author}{\bibfnamefont{X.~Y.} \bibnamefont{Zhao}} \bibnamefont{and}
  \bibinfo{author}{\bibfnamefont{D.}~\bibnamefont{Vanderbilt}},
  \bibinfo{journal}{Physical Review B} \textbf{\bibinfo{volume}{65}},
  \bibinfo{pages}{075105} (\bibinfo{year}{2002}).

\bibitem[{\citenamefont{Stapper et~al.}(1999)\citenamefont{Stapper, Bernasconi,
  Nicoloso, and Parrinello}}]{StapperBNP99}
\bibinfo{author}{\bibfnamefont{G.}~\bibnamefont{Stapper}},
  \bibinfo{author}{\bibfnamefont{M.}~\bibnamefont{Bernasconi}},
  \bibinfo{author}{\bibfnamefont{N.}~\bibnamefont{Nicoloso}}, \bibnamefont{and}
  \bibinfo{author}{\bibfnamefont{M.}~\bibnamefont{Parrinello}},
  \bibinfo{journal}{Physical Review B} \textbf{\bibinfo{volume}{59}},
  \bibinfo{pages}{797} (\bibinfo{year}{1999}).

\bibitem[{\citenamefont{Bogicevic and Wolverton}(2003)}]{BogicevicW03}
\bibinfo{author}{\bibfnamefont{A.}~\bibnamefont{Bogicevic}} \bibnamefont{and}
  \bibinfo{author}{\bibfnamefont{C.}~\bibnamefont{Wolverton}},
  \bibinfo{journal}{Physical Review B} \textbf{\bibinfo{volume}{67}},
  \bibinfo{pages}{024106} (\bibinfo{year}{2003}).

\bibitem[{\citenamefont{Fleming}(1977)}]{Fleming77}
\bibinfo{author}{\bibfnamefont{W.~J.} \bibnamefont{Fleming}},
  \bibinfo{journal}{Journal Of The Electrochemical Society}
  \textbf{\bibinfo{volume}{124}}, \bibinfo{pages}{21} (\bibinfo{year}{1977}).

\bibitem[{\citenamefont{Docquier and Candel}(2002)}]{DocquierC02}
\bibinfo{author}{\bibfnamefont{N.}~\bibnamefont{Docquier}} \bibnamefont{and}
  \bibinfo{author}{\bibfnamefont{S.}~\bibnamefont{Candel}},
  \bibinfo{journal}{Progress In Energy And Combustion Science}
  \textbf{\bibinfo{volume}{28}}, \bibinfo{pages}{107} (\bibinfo{year}{2002}).

\bibitem[{\citenamefont{Eichler}(2001)}]{Eichler01}
\bibinfo{author}{\bibfnamefont{A.}~\bibnamefont{Eichler}},
  \bibinfo{journal}{Physical Review B} \textbf{\bibinfo{volume}{6417}},
  \bibinfo{pages}{174103} (\bibinfo{year}{2001}).

\bibitem[{\citenamefont{Opeka et~al.}(2004)\citenamefont{Opeka, Talmy, and
  Zaykoski}}]{OpekaTZ04}
\bibinfo{author}{\bibfnamefont{M.~M.} \bibnamefont{Opeka}},
  \bibinfo{author}{\bibfnamefont{I.~G.} \bibnamefont{Talmy}}, \bibnamefont{and}
  \bibinfo{author}{\bibfnamefont{J.~A.} \bibnamefont{Zaykoski}},
  \bibinfo{journal}{Journal Of Materials Science}
  \textbf{\bibinfo{volume}{39}}, \bibinfo{pages}{5887} (\bibinfo{year}{2004}).

\bibitem[{\citenamefont{Monteverde and Bellosi}(2003)}]{MonteverdeB03}
\bibinfo{author}{\bibfnamefont{F.}~\bibnamefont{Monteverde}} \bibnamefont{and}
  \bibinfo{author}{\bibfnamefont{A.}~\bibnamefont{Bellosi}},
  \bibinfo{journal}{Journal Of The Electrochemical Society}
  \textbf{\bibinfo{volume}{150}}, \bibinfo{pages}{B552} (\bibinfo{year}{2003}).

\bibitem[{\citenamefont{Bongiorno et~al.}(2006)\citenamefont{Bongiorno, Forst,
  Kalia, Li, Marschall, Nakano, Opeka, Talmy, Vashishta, and
  Yip}}]{BongiornoFKLMOTVY06}
\bibinfo{author}{\bibfnamefont{A.}~\bibnamefont{Bongiorno}},
  \bibinfo{author}{\bibfnamefont{C.~J.} \bibnamefont{Forst}},
  \bibinfo{author}{\bibfnamefont{R.~K.} \bibnamefont{Kalia}},
  \bibinfo{author}{\bibfnamefont{J.}~\bibnamefont{Li}},
  \bibinfo{author}{\bibfnamefont{J.}~\bibnamefont{Marschall}},
  \bibinfo{author}{\bibfnamefont{A.}~\bibnamefont{Nakano}},
  \bibinfo{author}{\bibfnamefont{M.~M.} \bibnamefont{Opeka}},
  \bibinfo{author}{\bibfnamefont{I.~G.} \bibnamefont{Talmy}},
  \bibinfo{author}{\bibfnamefont{P.}~\bibnamefont{Vashishta}},
  \bibnamefont{and} \bibinfo{author}{\bibfnamefont{S.}~\bibnamefont{Yip}},
  \bibinfo{journal}{MRS Bulletin} \textbf{\bibinfo{volume}{31}},
  \bibinfo{pages}{410} (\bibinfo{year}{2006}).

\bibitem[{\citenamefont{Foster et~al.}(2001)\citenamefont{Foster, Sulimov,
  Gejo, Shluger, and Nieminen}}]{FosterSGSN01}
\bibinfo{author}{\bibfnamefont{A.~S.} \bibnamefont{Foster}},
  \bibinfo{author}{\bibfnamefont{V.~B.} \bibnamefont{Sulimov}},
  \bibinfo{author}{\bibfnamefont{F.~L.} \bibnamefont{Gejo}},
  \bibinfo{author}{\bibfnamefont{A.~L.} \bibnamefont{Shluger}},
  \bibnamefont{and} \bibinfo{author}{\bibfnamefont{R.~M.}
  \bibnamefont{Nieminen}}, \bibinfo{journal}{Physical Review B}
  \textbf{\bibinfo{volume}{64}}, \bibinfo{pages}{224108}
  (\bibinfo{year}{2001}).

\bibitem[{\citenamefont{Fiorentini and Gulleri}(2002)}]{FiorentiniG02}
\bibinfo{author}{\bibfnamefont{V.}~\bibnamefont{Fiorentini}} \bibnamefont{and}
  \bibinfo{author}{\bibfnamefont{G.}~\bibnamefont{Gulleri}},
  \bibinfo{journal}{Physical Review Letters} \textbf{\bibinfo{volume}{89}},
  \bibinfo{pages}{266101} (\bibinfo{year}{2002}).

\bibitem[{\citenamefont{Robertson et~al.}(2005)\citenamefont{Robertson, Xiong,
  and Falabretti}}]{RobertsonXF05}
\bibinfo{author}{\bibfnamefont{J.}~\bibnamefont{Robertson}},
  \bibinfo{author}{\bibfnamefont{K.}~\bibnamefont{Xiong}}, \bibnamefont{and}
  \bibinfo{author}{\bibfnamefont{B.}~\bibnamefont{Falabretti}},
  \bibinfo{journal}{Ieee Transactions On Device And Materials Reliability}
  \textbf{\bibinfo{volume}{5}}, \bibinfo{pages}{84} (\bibinfo{year}{2005}).

\bibitem[{\citenamefont{Grynszpan et~al.}(2007)\citenamefont{Grynszpan, Brauer,
  Anwand, Malaquin, Saud, Vickridge, and Briand}}]{GrynszpanBAMSVB07}
\bibinfo{author}{\bibfnamefont{R.~I.} \bibnamefont{Grynszpan}},
  \bibinfo{author}{\bibfnamefont{G.}~\bibnamefont{Brauer}},
  \bibinfo{author}{\bibfnamefont{W.}~\bibnamefont{Anwand}},
  \bibinfo{author}{\bibfnamefont{L.}~\bibnamefont{Malaquin}},
  \bibinfo{author}{\bibfnamefont{S.}~\bibnamefont{Saud}},
  \bibinfo{author}{\bibfnamefont{I.}~\bibnamefont{Vickridge}},
  \bibnamefont{and} \bibinfo{author}{\bibfnamefont{E.}~\bibnamefont{Briand}},
  \bibinfo{journal}{Nuclear Instruments \& Methods In Physics Research Section
  B-Beam Interactions With Materials And Atoms} \textbf{\bibinfo{volume}{261}},
  \bibinfo{pages}{888} (\bibinfo{year}{2007}).

\bibitem[{\citenamefont{Thome et~al.}(2007)\citenamefont{Thome, Gentils,
  Jagielski, Garrido, and Thome}}]{ThomeGJGT07}
\bibinfo{author}{\bibfnamefont{L.}~\bibnamefont{Thome}},
  \bibinfo{author}{\bibfnamefont{A.}~\bibnamefont{Gentils}},
  \bibinfo{author}{\bibfnamefont{J.}~\bibnamefont{Jagielski}},
  \bibinfo{author}{\bibfnamefont{F.}~\bibnamefont{Garrido}}, \bibnamefont{and}
  \bibinfo{author}{\bibfnamefont{T.}~\bibnamefont{Thome}},
  \bibinfo{journal}{Vacuum} \textbf{\bibinfo{volume}{81}},
  \bibinfo{pages}{1264} (\bibinfo{year}{2007}).

\bibitem[{\citenamefont{Thome et~al.}(2006)\citenamefont{Thome, Gentils,
  Jagielski, Garrido, and Thome}}]{ThomeGJGT06}
\bibinfo{author}{\bibfnamefont{L.}~\bibnamefont{Thome}},
  \bibinfo{author}{\bibfnamefont{A.}~\bibnamefont{Gentils}},
  \bibinfo{author}{\bibfnamefont{J.}~\bibnamefont{Jagielski}},
  \bibinfo{author}{\bibfnamefont{F.}~\bibnamefont{Garrido}}, \bibnamefont{and}
  \bibinfo{author}{\bibfnamefont{T.}~\bibnamefont{Thome}},
  \bibinfo{journal}{Nuclear Instruments \& Methods In Physics Research Section
  B-Beam Interactions With Materials And Atoms} \textbf{\bibinfo{volume}{250}},
  \bibinfo{pages}{106} (\bibinfo{year}{2006}).

\bibitem[{\citenamefont{Degueldre and Hellwig}(2003)}]{DegueldreH03}
\bibinfo{author}{\bibfnamefont{C.}~\bibnamefont{Degueldre}} \bibnamefont{and}
  \bibinfo{author}{\bibfnamefont{C.}~\bibnamefont{Hellwig}},
  \bibinfo{journal}{Journal Of Nuclear Materials}
  \textbf{\bibinfo{volume}{320}}, \bibinfo{pages}{96} (\bibinfo{year}{2003}).

\bibitem[{\citenamefont{Kresse and
  Furthmuller}(1996{\natexlab{a}})}]{KresseF96}
\bibinfo{author}{\bibfnamefont{G.}~\bibnamefont{Kresse}} \bibnamefont{and}
  \bibinfo{author}{\bibfnamefont{J.}~\bibnamefont{Furthmuller}},
  \bibinfo{journal}{Physical Review B} \textbf{\bibinfo{volume}{54}},
  \bibinfo{pages}{11169} (\bibinfo{year}{1996}{\natexlab{a}}).

\bibitem[{\citenamefont{Kresse and
  Furthmuller}(1996{\natexlab{b}})}]{KresseFJ96}
\bibinfo{author}{\bibfnamefont{G.}~\bibnamefont{Kresse}} \bibnamefont{and}
  \bibinfo{author}{\bibfnamefont{J.}~\bibnamefont{Furthmuller}},
  \bibinfo{journal}{Computational Materials Science}
  \textbf{\bibinfo{volume}{6}}, \bibinfo{pages}{15}
  (\bibinfo{year}{1996}{\natexlab{b}}).

\bibitem[{\citenamefont{Kresse and Joubert}(1999)}]{KresseJ99}
\bibinfo{author}{\bibfnamefont{G.}~\bibnamefont{Kresse}} \bibnamefont{and}
  \bibinfo{author}{\bibfnamefont{D.}~\bibnamefont{Joubert}},
  \bibinfo{journal}{Physical Review B} \textbf{\bibinfo{volume}{59}},
  \bibinfo{pages}{1758} (\bibinfo{year}{1999}).

\bibitem[{\citenamefont{Perdew et~al.}(1996)\citenamefont{Perdew, Burke, and
  Ernzerhof}}]{PerdewBE96}
\bibinfo{author}{\bibfnamefont{J.~P.} \bibnamefont{Perdew}},
  \bibinfo{author}{\bibfnamefont{K.}~\bibnamefont{Burke}}, \bibnamefont{and}
  \bibinfo{author}{\bibfnamefont{M.}~\bibnamefont{Ernzerhof}},
  \bibinfo{journal}{Physical Review Letters} \textbf{\bibinfo{volume}{77}},
  \bibinfo{pages}{3865} (\bibinfo{year}{1996}).

\bibitem[{\citenamefont{Jomard et~al.}(1999)\citenamefont{Jomard, Petit,
  Pasturel, Magaud, Kresse, and Hafner}}]{JomardPPMKH99}
\bibinfo{author}{\bibfnamefont{G.}~\bibnamefont{Jomard}},
  \bibinfo{author}{\bibfnamefont{T.}~\bibnamefont{Petit}},
  \bibinfo{author}{\bibfnamefont{A.}~\bibnamefont{Pasturel}},
  \bibinfo{author}{\bibfnamefont{L.}~\bibnamefont{Magaud}},
  \bibinfo{author}{\bibfnamefont{G.}~\bibnamefont{Kresse}}, \bibnamefont{and}
  \bibinfo{author}{\bibfnamefont{J.}~\bibnamefont{Hafner}},
  \bibinfo{journal}{Physical Review B} \textbf{\bibinfo{volume}{59}},
  \bibinfo{pages}{4044} (\bibinfo{year}{1999}).

\bibitem[{\citenamefont{H. et~al.}(2010)\citenamefont{H., I., P., and
  M}}]{JiangGRS10}
\bibinfo{author}{\bibfnamefont{J.}~\bibnamefont{H.}},
  \bibinfo{author}{\bibfnamefont{G.-A.~R.} \bibnamefont{I.}},
  \bibinfo{author}{\bibfnamefont{R.}~\bibnamefont{P.}}, \bibnamefont{and}
  \bibinfo{author}{\bibfnamefont{S.}~\bibnamefont{M}},
  \bibinfo{journal}{Physical Review B} \textbf{\bibinfo{volume}{81}},
  \bibinfo{pages}{085119} (\bibinfo{year}{2010}).

\bibitem[{\citenamefont{Makov and Payne}(1995)}]{MakovP95}
\bibinfo{author}{\bibfnamefont{G.}~\bibnamefont{Makov}} \bibnamefont{and}
  \bibinfo{author}{\bibfnamefont{M.~C.} \bibnamefont{Payne}},
  \bibinfo{journal}{Physical Review B} \textbf{\bibinfo{volume}{51}},
  \bibinfo{pages}{4014} (\bibinfo{year}{1995}).

\bibitem[{\citenamefont{McQuarrie}(1976)}]{McQuarrieDA}
\bibinfo{author}{\bibfnamefont{D.~A.} \bibnamefont{McQuarrie}},
  \bibinfo{journal}{Statistical Mechanics, Harper Collins Publishers}
  (\bibinfo{year}{1976}).

\bibitem[{\citenamefont{Villars}(1929)}]{Villars29}
\bibinfo{author}{\bibfnamefont{D.~S.} \bibnamefont{Villars}},
  \bibinfo{journal}{Proceedings of the National Academy of Sciences}
  \textbf{\bibinfo{volume}{15}}, \bibinfo{pages}{705} (\bibinfo{year}{1929}).

\bibitem[{\citenamefont{Foster et~al.}(2002{\natexlab{a}})\citenamefont{Foster,
  Shluger, and Nieminen}}]{FosterSN02}
\bibinfo{author}{\bibfnamefont{A.~S.} \bibnamefont{Foster}},
  \bibinfo{author}{\bibfnamefont{A.~L.} \bibnamefont{Shluger}},
  \bibnamefont{and} \bibinfo{author}{\bibfnamefont{R.~M.}
  \bibnamefont{Nieminen}}, \bibinfo{journal}{Physical Review Letters}
  \textbf{\bibinfo{volume}{89}}, \bibinfo{pages}{225901}
  (\bibinfo{year}{2002}{\natexlab{a}}).

\bibitem[{\citenamefont{Kralik et~al.}(1998)\citenamefont{Kralik, Chang, and
  Louie}}]{KralikCL98}
\bibinfo{author}{\bibfnamefont{B.}~\bibnamefont{Kralik}},
  \bibinfo{author}{\bibfnamefont{E.~K.} \bibnamefont{Chang}}, \bibnamefont{and}
  \bibinfo{author}{\bibfnamefont{S.~G.} \bibnamefont{Louie}},
  \bibinfo{journal}{Physical Review B} \textbf{\bibinfo{volume}{57}},
  \bibinfo{pages}{7027} (\bibinfo{year}{1998}).

\bibitem[{\citenamefont{Fabris et~al.}(2002)\citenamefont{Fabris, Paxton, and
  Finnis}}]{FabrisPF02}
\bibinfo{author}{\bibfnamefont{S.}~\bibnamefont{Fabris}},
  \bibinfo{author}{\bibfnamefont{A.~T.} \bibnamefont{Paxton}},
  \bibnamefont{and} \bibinfo{author}{\bibfnamefont{M.~W.}
  \bibnamefont{Finnis}}, \bibinfo{journal}{Acta Materialia}
  \textbf{\bibinfo{volume}{50}}, \bibinfo{pages}{005171}
  (\bibinfo{year}{2002}).

\bibitem[{\citenamefont{Kroger and Vink}(1957)}]{KrogerV57}
\bibinfo{author}{\bibfnamefont{F.~A.} \bibnamefont{Kroger}} \bibnamefont{and}
  \bibinfo{author}{\bibfnamefont{H.~J.} \bibnamefont{Vink}},
  \bibinfo{journal}{Solid State Physics - Advances in Research and
  Applications, Academic Press, New York}  (\bibinfo{year}{1957}).

\bibitem[{\citenamefont{Blochl and Stathis}(1999)}]{BlochlS99}
\bibinfo{author}{\bibfnamefont{P.~E.} \bibnamefont{Blochl}} \bibnamefont{and}
  \bibinfo{author}{\bibfnamefont{H.}~\bibnamefont{Stathis}},
  \bibinfo{journal}{Physical Review Letters} \textbf{\bibinfo{volume}{83}},
  \bibinfo{pages}{372} (\bibinfo{year}{1999}).

\bibitem[{\citenamefont{Tian}(2004)}]{Tian04}
\bibinfo{author}{\bibfnamefont{S.~X.} \bibnamefont{Tian}},
  \bibinfo{journal}{Journal Of Physical Chemistry B}
  \textbf{\bibinfo{volume}{108}}, \bibinfo{pages}{20388}
  (\bibinfo{year}{2004}).

\bibitem[{\citenamefont{Avdeev et~al.}(1997)\citenamefont{Avdeev, Ruzankin, and
  Zhidomirov}}]{AvdeevRZ97}
\bibinfo{author}{\bibfnamefont{V.~I.} \bibnamefont{Avdeev}},
  \bibinfo{author}{\bibfnamefont{S.~F.} \bibnamefont{Ruzankin}},
  \bibnamefont{and} \bibinfo{author}{\bibfnamefont{G.~M.}
  \bibnamefont{Zhidomirov}}, \bibinfo{journal}{Journal Of Structural Chemistry}
  \textbf{\bibinfo{volume}{38}}, \bibinfo{pages}{519} (\bibinfo{year}{1997}).

\bibitem[{\citenamefont{Foster et~al.}(2002{\natexlab{b}})\citenamefont{Foster,
  Gejo, Shluger, and Nieminen}}]{FosterGSN02}
\bibinfo{author}{\bibfnamefont{A.~S.} \bibnamefont{Foster}},
  \bibinfo{author}{\bibfnamefont{F.~L.} \bibnamefont{Gejo}},
  \bibinfo{author}{\bibfnamefont{A.~L.} \bibnamefont{Shluger}},
  \bibnamefont{and} \bibinfo{author}{\bibfnamefont{R.~M.}
  \bibnamefont{Nieminen}}, \bibinfo{journal}{Physical Review B}
  \textbf{\bibinfo{volume}{65}}, \bibinfo{pages}{174117}
  (\bibinfo{year}{2002}{\natexlab{b}}).

\bibitem[{\citenamefont{Huggins}(2001)}]{Huggins01}
\bibinfo{author}{\bibfnamefont{R.~A.} \bibnamefont{Huggins}},
  \bibinfo{journal}{Solid State Ionics} \textbf{\bibinfo{volume}{143}},
  \bibinfo{pages}{3} (\bibinfo{year}{2001}).

\bibitem[{\citenamefont{Bogdanov et~al.}(1998)\citenamefont{Bogdanov, Dimitrov,
  Lutskaya, and Tairov}}]{BogdanovDLT98}
\bibinfo{author}{\bibfnamefont{K.~P.} \bibnamefont{Bogdanov}},
  \bibinfo{author}{\bibfnamefont{D.~T.} \bibnamefont{Dimitrov}},
  \bibinfo{author}{\bibfnamefont{O.~F.} \bibnamefont{Lutskaya}},
  \bibnamefont{and} \bibinfo{author}{\bibfnamefont{Y.~M.}
  \bibnamefont{Tairov}}, \bibinfo{journal}{Semiconductors}
  \textbf{\bibinfo{volume}{32}}, \bibinfo{pages}{1033} (\bibinfo{year}{1998}).

\bibitem[{\citenamefont{Barsoum}(1997)}]{Barsoum97}
\bibinfo{author}{\bibfnamefont{M.~W.} \bibnamefont{Barsoum}},
  \bibinfo{journal}{Fundamentals of ceramics, Institute of Physics Publishing}
  (\bibinfo{year}{1997}).

\bibitem[{\citenamefont{Zheng et~al.}(2007)\citenamefont{Zheng, Ceder, Maxisch,
  Chim, and Choi}}]{ZhengCMCC07}
\bibinfo{author}{\bibfnamefont{J.~X.} \bibnamefont{Zheng}},
  \bibinfo{author}{\bibfnamefont{G.}~\bibnamefont{Ceder}},
  \bibinfo{author}{\bibfnamefont{T.}~\bibnamefont{Maxisch}},
  \bibinfo{author}{\bibfnamefont{W.~K.} \bibnamefont{Chim}}, \bibnamefont{and}
  \bibinfo{author}{\bibfnamefont{W.~K.} \bibnamefont{Choi}},
  \bibinfo{journal}{Physical Review B} \textbf{\bibinfo{volume}{75}},
  \bibinfo{pages}{104112} (\bibinfo{year}{2007}).

\bibitem[{\citenamefont{Brouwer}(1954)}]{Brouwer54}
\bibinfo{author}{\bibfnamefont{G.}~\bibnamefont{Brouwer}},
  \bibinfo{journal}{Philips Research Reports} \textbf{\bibinfo{volume}{9}},
  \bibinfo{pages}{366} (\bibinfo{year}{1954}).

\bibitem[{\citenamefont{Ackermann et~al.}(1977)\citenamefont{Ackermann, Garg,
  and Rauh}}]{AckermannGR77}
\bibinfo{author}{\bibfnamefont{R.~J.} \bibnamefont{Ackermann}},
  \bibinfo{author}{\bibfnamefont{S.~P.} \bibnamefont{Garg}}, \bibnamefont{and}
  \bibinfo{author}{\bibfnamefont{E.~G.} \bibnamefont{Rauh}},
  \bibinfo{journal}{Journal of The American Ceramic Society}
  \textbf{\bibinfo{volume}{60}}, \bibinfo{pages}{341} (\bibinfo{year}{1977}).

\bibitem[{\citenamefont{Pruneda et~al.}(2004)\citenamefont{Pruneda, Archer, and
  Artacho}}]{PrunedaAA04}
\bibinfo{author}{\bibfnamefont{J.~M.} \bibnamefont{Pruneda}},
  \bibinfo{author}{\bibfnamefont{T.~D.} \bibnamefont{Archer}},
  \bibnamefont{and} \bibinfo{author}{\bibfnamefont{E.}~\bibnamefont{Artacho}},
  \bibinfo{journal}{Physical Review B} \textbf{\bibinfo{volume}{70}},
  \bibinfo{pages}{104111} (\bibinfo{year}{2004}).

\end{thebibliography}

 \clearpage
 
 \clearpage

 \clearpage

 \clearpage
 \begin{figure}[thbp]
   \centering
   \includegraphics[width=0.5\textheight]{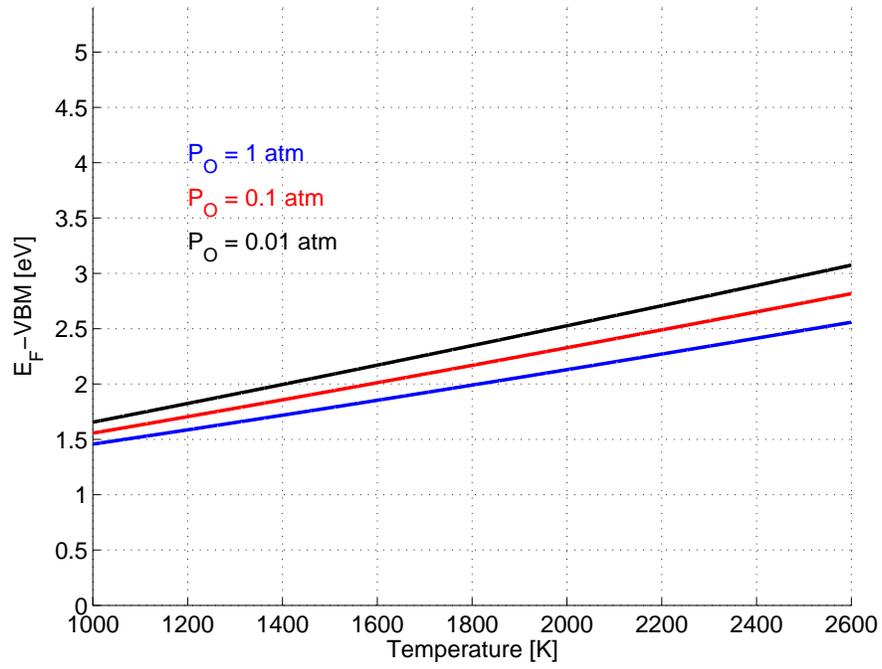}
   \caption{ The chemical potential (Fermi energy) of an undoped cubic ZrO$_{2}$
     system as a function of temperature at different oxygen partial pressures.}
   \label{FermiEnergySystem}
 \end{figure}

\end{document}